\documentclass[aps,prb,twocolumn,amsmath,amssymb,groupedaddress,floatfix,longbibliography]{revtex4-1}
\setcitestyle{square,numbers}
\usepackage{graphicx}  
\usepackage{bm}        
\usepackage{amssymb}   
\usepackage{amsmath}
\usepackage{color}
\usepackage{epstopdf}

\usepackage[plainpages=false,pdfpagelabels,colorlinks=true,linkcolor=red,urlcolor=blue,citecolor=blue,pdftitle={Title},pdfauthor={},pdfdisplaydoctitle=true,pdfduplex=DuplexFlipLongEdge]{hyperref}

\begin{document}

\title{Grand-canonical Peierls theory for atomic wires on substrates}
\author{Yasemin Erg\"{u}n}
\email[E-mail: ]{yasemin.erguen@itp.uni-hannover.de}
\author{Eric Jeckelmann}
\email[E-mail: ]{eric.jeckelmann@itp.uni-hannover.de}
\affiliation{Leibniz Universit\"{a}t Hannover, Institut f\"{u}r Theoretische Physik, Appelstr.~2, 30167 Hannover, Germany}

\date{Draft of \today}

\begin{abstract}
We present a generic grand-canonical theory for the Peierls transition
in atomic wires deposited on semiconducting substrates such as In/Si(111)
using a mean-field solution of the one-dimensional Su-Schrieffer-Heeger model.
We show that this simple low-energy effective model for atomic wires can explain naturally
the occurrence of a first-order Peierls transition between a uniform metallic
phase at high-temperature and a dimerized insulating phase at low temperature
as well as the existence of a metastable uniform state below the critical temperature.
\end{abstract}

\maketitle

\section{Introduction}

The Peierls instability of a one-dimensional metal coupled to lattice vibrations~\cite{Peierls}
is a hallmark of low-dimensional physics.
It is known to play an important role in various strongly anisotropic bulk materials
and there is a well-established theoretical framework to describe these quasi-one-dimensional
systems at a fixed electronic density~\cite{Gruener00,heeg88,baeriswyl92,solyom3}.
In particular, the canonical Peierls theory predicts a continuous transition from a metallic state with a uniform lattice at high temperature
to an insulating state with a distorted lattice at low temperature.

Atomic wires deposited on a semiconducting substrate~\cite{Oncel08,Snijders10}  represent another possible
realization of one-dimensional electronic systems in which the Peierls instability
could play a role.
Indeed, the Peierls mechanism
was invoked to explain the metal-insulator transition accompanied by a structural transition
with a doubling of the unit cell observed in indium wires on a Si(111) surface~\cite{yeom99}.
Various other explanations have been proposed for this transition, however, and the 
relevance of the Peierls mechanism remains controversial.
The common problem with most previous interpretations of experiments and first-principles simulations
 is that they do not consider
how the semiconducting substrate modifies the predictions of the one-dimensional Peierls theory.

Recent experimental evidence and first-principles simulations for In/Si(111)  suggest that the 
metallic uniform state remains metastable below the critical temperature and that the transition 
is first order in this system~\cite{hatt11,wall12,schm12,klas14,zhan14,Jeckelmann16,spei16,hatt17}.
In Ref.~\cite{Jeckelmann16} it was shown that the first-principles simulation results, the metastability
of the metallic uniform state, and the first-order transition could be understood within a grand-canonical 
Peierls theory based on an effective one-dimensional low-energy model. 
The indium wires were described  by a specifically constructed 
Hamiltonian of the Su-Schrieffer-Heeger (SSH) type~\cite{su79,su80}
including four electronic bands and three commensurate lattice distortion modes
while the substrate was treated  as a charge reservoir for the wire subsystem.
However, the first-order transition was only obtained under the assumption that the
chemical potential must vary with temperature to preserve the overall charge neutrality.
Moreover, the model was investigated purely numerically and only for the parameters that were obtained from
first-principles simulations of In/Si(111).

In this paper we present a generic grand-canonical Peierls theory based on  the original SSH model 
with one electronic band and one commensurate lattice distortion mode in the mean-field approximation (dimerization).
The substrate only acts as an electron reservoir and sets the chemical potential for the wires.
Using only analytical  results and basic numerical calculations
we demonstrate that in a grand-canonical Peierls system the high-temperature uniform metallic state can remain 
thermodynamically metastable below the critical temperature. Additionally,
we show that the structural Peierls phase transition 
can be first order as a function of temperature for a fixed chemical potential. 
Moreover, we find that the metal-insulator
transition is always first order.

In the next section we introduce the grand-canonical mean-field approach for atomic wires on semiconducting substrates
based on the SSH model. In Sec.~\ref{sec:half} we recapitulate the known results
for the Peierls transition in the SSH model at half filling.
Our results for the SSH model in the grand-canonical ensemble are presented in Sec.~\ref{sec:gc}.
Finally, Sec.~\ref{sec:conclusion} contains our conclusions.
Some details are presented in two appendices

\section{Grand-canonical model for the Peierls transition \label{sec:model}}

\subsection{SSH model \label{sec:ssh}}

\begin{figure}
\includegraphics[width=0.9\columnwidth]{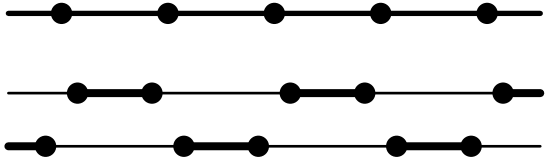}
\caption{\label{fig:ssh}
Schematic of the uniform lattice configuration (top) and the two dimerized configurations (bottom) in the SSH model.
The disks represent the sites (atoms) and the line widths show the strength of the bond order or hopping term.
}
\end{figure}

The SSH model~\cite{su79,su80,heeg88,baeriswyl92} 
is the standard model for charge density waves (CDW) on bonds caused by a Peierls distortion of bond lengths.
It consists of a one-dimensional lattice  with $L$ sites (ions) and $N/2$ electrons of each spin.
The lattice degrees of freedom are treated classically.
The position of site $j$ along the lattice axis
is given by $x_j = j a + u_j$ where $a$ is the lattice constant of the
uniform lattice configuration ($x_j=j a$) shown in Fig.~\ref{fig:ssh}, while $u_j$ describes the deviation from 
this position.
The Hamilton function for the lattice degrees of freedom (without coupling to the electrons)
is
\begin{equation}
\label{eq:lattice}
{\cal H} = \sum_j \frac{P_j^2}{2M} + V  
\end{equation}
where $P_j$ designates the conjugate momentum of $u_j$, $M$ is the effective mass of the
ions, and
$V$ is the lattice elastic energy.
For small deviations $u_j$ the lattice elastic energy can be approximated
by a harmonic potential 
\begin{equation}
\label{eq:potential}
V = \frac{K}{2} \sum_j (u_{j+1} - u_j)^2 
\end{equation}
with the spring constant $K$.
We will use periodic boundary conditions $u_{L+1} \equiv u_{1}$ and thus 
sums over the site index $j$ always runs from 1 to $L$.

A tight-binding Hamiltonian is used for the electronic degrees of freedom and 
it is assumed
that the only relevant hopping terms are between nearest-neighbor sites
(including the coupling to the lattice deformations).
The Hamilton operator is  
\begin{equation}
\label{eq:hamiltonian}
H = - \sum_{j, \sigma} t_{j} \left ( c^{\dag}_{j,\sigma}
c^{\phantom{\dag}}_{j+1,\sigma} + c^{\dag}_{j+1,\sigma} c^{\phantom{\dag}}_{j,\sigma} \right )
\end{equation}
where the operator $c^{\dag}_{j,\sigma}$ ($c^{\phantom{\dag}}_{j,\sigma}$) creates
(annihilates) an electron with spin $\sigma (=\uparrow,\downarrow)$ on site $j$.  
The hopping term $t_{j}$ depends on the distance between the sites $j$ and $j+1$,
i.e. $t_j = t(u_{j+1}-u_{j})$. In the SSH approach a linear dependence is assumed
for small deviations $u_{j}$
\begin{equation}
\label{eq:hopping}  
t_j = t_0 - \alpha (u_{j+1}-u_{j})
\end{equation}
with the electron-phonon coupling constant $\alpha \geq 0$ 
and the value $t_0$ of the hopping for the uniform lattice configuration.
As the sign of the hopping terms can be changed with a simple gauge transformation
$c^{\dag}_{j\sigma} \rightarrow (-1)^j c^{\dag}_{j\sigma}$,
it is sufficient to consider the case $t_j > 0 \ (\Rightarrow t_0>0)$.

Note that this linear approximation for $t(u_{j+1}-u_{j})$
is inconsistent with the quadratic approximation for the lattice potential~(\ref{eq:potential}).
As this inconsistency does not change the results qualitatively, we will
keep to the SSH choice in this paper. See the appendix~\ref{app:hopping} for more details.

\subsection{Dimerization \label{sec:mf}}

In this work we will exclusively consider the mean-field approximation for a commensurate Peierls distortion of
period two (dimerization) and thus assume that
\begin{equation}
\label{eq:mf}
u_j = (-1)^j u
\end{equation}
with a constant alternating site displacement $u$~\cite{su79,su80,heeg88,baeriswyl92}.
This corresponds to alternating long and short bonds, as shown in Fig.~\ref{fig:ssh}.
This particular lattice distortion is the normal mode of the classical oscillator system~(\ref{eq:lattice}) 
with wave number $Q=\pi/a$ and thus the unstable $2k_{\text{F}}$-mode of the Peierls theory at half filling
[$k_{\text{F}}=\pi/(2a) \Rightarrow Q = \pi/a$].
The displacement amplitude $u$ is related to the usual Peierls order parameter (or dimerization parameter) by
\begin{equation}
\label{eq:Delta}
\Delta = 4 \alpha u 
\end{equation}
It will be convenient to work with the dimensionless order parameter
\begin{equation}
\delta = \frac{\Delta}{2 t_0} = \frac{2 \alpha u}{t_0} . 
\end{equation}
The lattice elastic energy~(\ref{eq:potential}) becomes 
\begin{equation}
V = \frac{L t_0}{\pi \lambda} \delta^2 
\end{equation}
with the dimensionless SSH electron-phonon coupling 
\begin{equation}
\lambda = \frac{2 \alpha^2}{\pi K t_0}.
\end{equation}
In Ref.~\cite{Jeckelmann16} it was shown that $\lambda = 0.37$ and $0.18$
were realistic values for the shear and rotary Peierls modes
of indium wires on Si(111), respectively.
The lattice kinetic energy can also be written as a function of the order parameter
\begin{equation}
\label{eq:kinetic}
E_{\text{K}} = \frac{L t_0}{\pi \lambda} \frac{\dot{\delta}^2}{\omega_0^2}  
\end{equation}
where $\dot{\delta}$ designates the time derivative of the order parameter $\delta$ and
\begin{equation}
\label{eq:bare_phonon}
\omega_0 = \sqrt{\frac{4K}{M}} 
\end{equation}
is the bare frequency of amplitude fluctuations of the order parameter (i.e., the normal mode with wave number $Q=\pi/a$)
when the lattice is not coupled to the electrons.

\begin{figure}
\includegraphics[width=0.9\columnwidth]{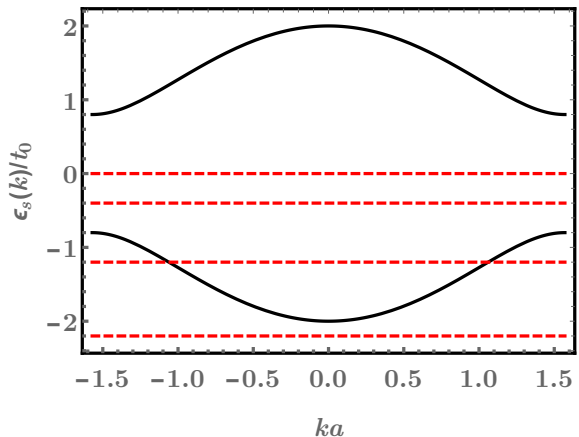}
\caption{\label{fig:dispersion}
Dispersion~(\ref{eq:dispersion}) of the single-electron eigenenergies for $\Delta=2\delta t_0= 0.8t_0$.
The horizontal dashed lines show the four different relative positions of the chemical
potential: $\mu=0$ (half filling), $0 > \mu > -\Delta$,
$-\Delta > \mu > -2 t_0$, and $-2t_0 > \mu$ (see the discussion in Sec.~\ref{sec:gs}).
}
\end{figure}

From the assumption~(\ref{eq:mf}) follows that  the hopping terms $t_j$ alternate
between two values and thus
we can compute the single-particle eigenstates
of the electronic Hamiltonian~(\ref{eq:hamiltonian}) exactly.
The single-electron eigenenergies are given by
\begin{equation}
\label{eq:dispersion}
\varepsilon_{s}(k)  =  s 
2 t_0 \sqrt{ \cos^2(ka)  + \delta^2 \sin^2(ka)}
\end{equation}
with the band index $s=\pm 1$ and the wave number $k = 2 \pi z/(La)$ for integers $-L/4 < z \leq L/4$, which implies
$ k \in (-\pi/(2a),\pi/(2a)]$.
We see in Fig.~\ref{fig:dispersion} that the spectrum contains a gap of width 
\begin{equation}
\label{eq:gap}
E_{\text{g}} = 2 \vert \Delta \vert = 4 t_0 \vert \delta \vert
\end{equation}
centered around the energy $\varepsilon = 0$. Consequently, $\delta$ (or $\Delta$) is also called  gap
parameter. At half filling (and more generally
for a chemical potential $\vert \mu \vert < \vert \Delta \vert = 2 t_0 \vert \delta \vert$) 
the electronic  system is an insulator for $\delta \neq 0$ while it is metallic for 
$\delta = 0$.
The assumption of small position deviations $\vert u_j \vert = \vert u \vert$ used above can
now be formulated quantitatively as the condition $\vert \delta \vert \ll  1$.

The bond order between sites $j$ and $j+1$ is defined by
\begin{equation}
\label{eq:bondorder}
P_{j} = \frac{1}{2} \sum_{\sigma} 
\left \langle c^{\dag}_{j+1,\sigma}
c^{\phantom{\dag}}_{j,\sigma} + c^{\dag}_{j+1,\sigma} c^{\phantom{\dag}}_{j,\sigma}
\right \rangle
\end{equation}
where $\langle \dots \rangle$ is the expectation value for the ground state or the appropriate
statistical operator representing the electronic degrees of freedom.
This quantity is proportional to the electronic density on the bond between sites $j$ and $j+1$.
It is constant for a uniform lattice but oscillates
between two values for a dimerized lattice. 
Thus a vanishing order parameter $\delta = 0$ corresponds to a uniform
chain while a finite order parameter $\delta \neq 0$ corresponds to a CDW on the bonds
and a dimerized lattice.  

Moreover, in the canonical ensemble at half filling,
the uniform configuration yields a metallic state while the dimerized one
corresponds to an insulator as discussed above.
As the SSH model in the mean-field approximation~(\ref{eq:mf}) is invariant 
under simultaneous reflection ($u \rightarrow -u$) and  translation ($j \rightarrow j+1$) transformations,
configurations with opposite order parameters ($\delta$ and $-\delta$) have the same energy.
Consequently, the ground state is doubly degenerate in the dimerized phase, as illustrated in 
Fig.~\ref{fig:ssh}.

\subsection{Grand-canonical potential}

Our goal is to determine the equilibrium properties of the SSH model at finite temperature $T$.
For wire-substrate systems such as In/Si(111), the number of electrons $N$ is not fixed but the chemical potential $\mu$ 
is set by the substrate, which acts as a reservoir~\cite{Jeckelmann16}.
Thus we will use the grand-canonical ensemble for the electronic degrees of freedom 
but the canonical ensemble for the lattice degrees of freedom because we assume that 
the number of sites (atoms) in the chain is fixed.

After tracing out the electronic degrees of freedom we obtain the
grand-canonical potential of the full system per lattice site 
\begin{equation}
\label{eq:free_energy}
\phi = \frac{t_0}{\pi \lambda} \delta^2 -
\frac{k_{\text{B}} T}{L} \sum_{k,s,\sigma} \ln\left (1 + \exp\left (-\beta \left [\varepsilon_s(k) - \mu\right ]\right ) \right)
\end{equation}
with $1/\beta = k_{\text{B}} T$.
In the thermodynamic limit $L\rightarrow \infty$ one can write
\begin{eqnarray}
\label{eq:free_energy2}
\phi & = & \frac{t_0}{\pi \lambda} \delta^2 -\mu - k_{\text{B}} T \ln(2)  \\
&&
-k_{\text{B}} T \int_{\vert \Delta \vert}^{2 t_0} d\varepsilon  D(\varepsilon) \ln \left ( \cosh(\beta \mu) + \cosh(\beta \varepsilon) \right )
\nonumber
\end{eqnarray}
where the single-particle density of states is given by 
\begin{equation}
D(\varepsilon) =\frac{2}{\pi} \frac{\vert \varepsilon \vert }{\sqrt{(4t_0^2 -\varepsilon^2)(\varepsilon^2 - \Delta^2)}} 
\end{equation}
for $\Delta < \vert \epsilon \vert  < 2t_0$. 
Using a substitution $z=\varepsilon/(2t_0)$ one can easily verify
that $\phi$ depends only on the two thermodynamical variables $T$ and $\mu$,
the two model parameters $\lambda$ and $t_0$ (the latter just sets the energy scale),
and the order parameter $\delta$.
We also note that $\phi$ is an even function of $\delta$.

Within this mean-field, semi-classical approach 
the grand-canonical potential~(\ref{eq:free_energy2}) plays the role of the Landau's free energy
for the order parameter $\delta$ of the commensurate Peierls transition~\cite{Gruener00}.
The actual grand-canonical potential $\phi(T,\mu)$ and the stable configurations $\delta(T,\mu)$
are given by the minima of~(\ref{eq:free_energy2}) with respect to variations of $\delta$.
Thus our main goal is to determine the stable configurations and the related observables
as a function of $T$ and $\mu$, as well as the single remaining model parameter $\lambda$.

In the grand-canonical ensemble the average electronic density 
for a given temperature $T$ and chemical potential $\mu$ is given by 
\begin{equation}
\label{eq:density}
\rho = \frac{N}{L} = \frac{1}{L} \sum_{k,s,\sigma}  f(\varepsilon_s(k))
\end{equation}
with the Fermi-Dirac distribution
\begin{equation}
f(\varepsilon) = \frac{1}{1+\exp(\beta(\varepsilon-\mu))} .
\end{equation}
In the thermodynamic limit we can write
\begin{equation}
\label{eq:density2}
\rho = 1 + \int_{\vert \Delta \vert}^{2 t_0} d\varepsilon  D(\varepsilon) 
\frac{\sinh(\beta \mu)}{\cosh(\beta \mu) + \cosh(\beta \varepsilon) }. 
\end{equation}
We see that for $\mu=0$ the electronic band is half filled ($\rho=1$) while
less than half filling ($0 \leq \rho < 1)$ corresponds to $\mu < 0$ and 
more than half filling ($1 < \rho \leq 2)$ to $\mu > 0$.
Because the SSH model is invariant under the particle-hole transformation 
$c^{\dag}_{j\sigma} \rightarrow (-1)^j c^{\phantom{\dag}}_{j\sigma}$, 
the results are similar for $\mu \leq 0$ and $\mu \geq 0$ , and thus
we will discuss the first case only.

Eqs.~(\ref{eq:free_energy2}) and (\ref{eq:density2}) are the starting point
for studying thermodynamical Peierls transitions in the SSH model at the mean-field level.
Usually, it is assumed that the electronic density is fixed and the grand-canonical ensemble
is used only for computational convenience~\cite{Gruener00,solyom3}. 
Therefore, the value of the chemical potential is set
by  Eq.~(\ref{eq:density2}) for the desired value of $\rho$.
In Sec.~\ref{sec:half} we will summarize the results obtained with this assumption for the Peierls transition
at half filling.
In Sec.~\ref{sec:gc} we will then generalize these results for a fixed chemical potential.

\section{Results at fixed band filling \label{sec:half}}

The conventional Peierls theory assumes a fixed electronic density $\rho$. 
For the dimerized SSH model the band is usually half filled ($\rho=1$)
which corresponds to $\mu=0$ according to~(\ref{eq:density2}).
Here we summarize the most important results for this case~\cite{su79,su80,heeg88,baeriswyl92,Gruener00,solyom2,solyom3}.
The grand-canonical potential~(\ref{eq:free_energy2}) is then simplified
\begin{eqnarray}
\label{eq:free_energy3}
\phi & = & \frac{t_0}{\pi \lambda} \delta^2 - k_{\text{B}} T \ln(2)  \\
&&
-k_{\text{B}} T \int_{\vert \Delta \vert}^{2 t_0} d\varepsilon  D(\varepsilon) \ln \left ( 1 + \cosh(\beta \varepsilon) \right ).
\nonumber
\end{eqnarray}
For high temperatures ($k_{\text{B}} T \gg t_0$) it can be approximated by
\begin{equation}
\phi = \frac{t_0}{\pi \lambda} \delta^2 - k_{\text{B}} T \ln(4)  . 
\end{equation}
Therefore, the uniform metallic configuration $(\delta =0)$ 
is the only stable phase at high temperatures.

\begin{figure}
\includegraphics[width=0.9\columnwidth]{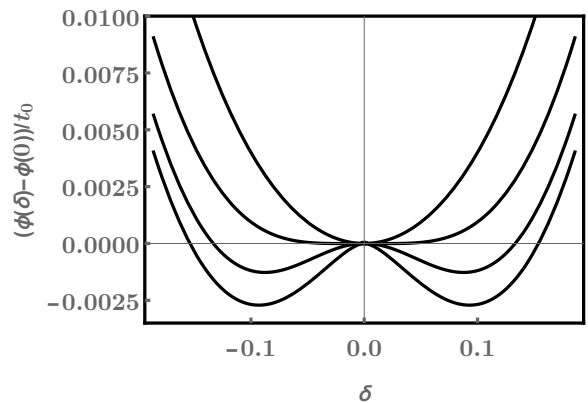}
\caption{\label{fig:potential} 
Grand-canonical potential~(\ref{eq:free_energy2}) at $\mu=0$ (half filling) [see Eq.~(\ref{eq:free_energy3})]  as a function 
of the order parameter $\delta$ for $\lambda = 0.18$ and four different temperatures (from bottom to top):
$T\approx 0$, $T\approx T_{\text{P}}/2$, $T\approx T_{\text{P}}$, and $T \approx 2 T_{\text{P}}$.
}
\end{figure}

For low temperatures ($T \rightarrow 0$) the grand-canonical potential~(\ref{eq:free_energy3}) yields
the ground-state energy
\begin{eqnarray}
\label{eq:gs_energy}
\phi & = & \frac{t_0}{\pi \lambda} \delta^2 - \int_{\vert \Delta \vert}^{2 t_0} d\varepsilon  D(\varepsilon) 
\varepsilon  \nonumber \\
& = &  \frac{t_0}{\pi \lambda} \delta^2 - \frac{4t_0}{\pi} E(1-\delta^2),
\end{eqnarray}
where $E(m)$ is the complete elliptic integral of the second kind.
It is well known that this expression is a double well potential
as a function of $\delta$ ($\vert \delta \vert < 1$) with
two minima at $\delta = \pm \delta_0 \neq 0$
and a local maximum at $\delta=0$, see the lower curve in Fig.~\ref{fig:potential}.
Throughout this paper we denote with $\delta_0 = \Delta_0/(2t_0)=\delta(T=0,\mu=0)$ the absolute value of the
order parameter for the stable configuration at half filling and zero temperature.
 
For small $\delta$ we can expand~(\ref{eq:gs_energy}) up to the lowest relevant order in $\delta$
\begin{equation}
\label{eq:expansion}
\phi =\frac{t_0}{\pi \lambda} \delta^2 - \frac{4t_0}{\pi} \left [ 1 +\frac{1}{2} \left ( 
\ln \left (\frac{4}{\vert\delta\vert}\right ) - \frac{1}{2} \right ) \delta^2 \right ] .
\end{equation}
Then one can easily verify that $\phi$ has a double minimum for
$\delta = \pm \delta_0$ with
\begin{equation}
\label{eq:delta0}
\delta_0 = \frac{4}{e} \exp \left (-\frac{1}{2\lambda} \right ).
\end{equation}
We see that this calculation is valid for weak electron-phonon coupling because
$\delta_0 \ll 1 \Leftrightarrow \lambda \ll 1$.
As an example, for $\lambda=0.18$ we obtain $\delta_0 \approx 0.0915$ from~(\ref{eq:delta0})
and  $\delta_0 \approx 0.0930$ from the numerical minimization of the energy~(\ref{eq:gs_energy}).
The condensation energy of the dimerized state per site (i.e., the difference between
the ground-state energy~(\ref{eq:expansion}) at $\delta=0$ and at its minimum) is 
\begin{equation}
\label{eq:condensation}
\Delta\phi = \frac{t_0}{\pi} \delta_0^2.
\end{equation}

The Peierls theory predicts the existence of Raman-active collective excitations
(electronic CDW and lattice vibrations),
which correspond to amplitude oscillations of the order parameter $\delta$ around its
equilibrium configuration in the mean-field approach~\cite{schu78,horo78,tuti91,Gruener00}.
The effective spring constant for these oscillations is given by 
the second derivative of $\phi$ with respect to $\delta$ at its minimum.
Combined with the kinetic energy~(\ref{eq:kinetic}) this leads to
the renormalized (phonon) frequency for amplitude fluctuations
\begin{equation}
\label{eq:omega}
\Omega^2 = \frac{\pi \lambda}{2 t_0} \frac{d^2 \phi}{d\delta^2} \omega_0^2
\end{equation}
where $\omega_0$ is the frequency~(\ref{eq:bare_phonon}) of the bare phonon mode with wave number $Q=2k_{\text{F}}$.
For the weak-coupling regime ($\lambda \ll 1$) 
we obtain using~(\ref{eq:expansion}) and~(\ref{eq:delta0})~\cite{heeg88,baeriswyl92,horo82} 
\begin{equation}
\label{eq:omega0}
\Omega^2 = 2 \lambda \omega_0^2.
\end{equation}

\begin{figure}
\includegraphics[width=0.9\columnwidth]{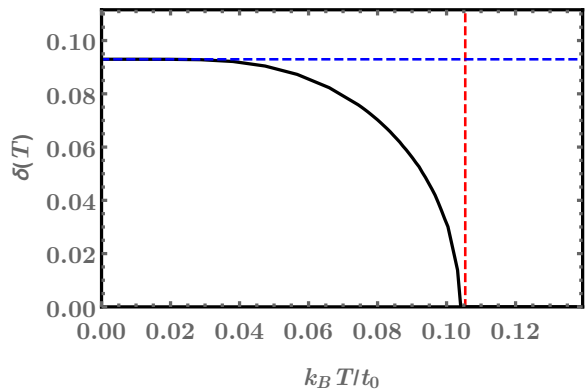}
\caption{\label{fig:delta} 
Order parameter $\delta$ as a function of temperature at half filling 
for the electron-phonon coupling $\lambda = 0.18$.  
The horizontal line shows the zero-temperature value $\delta_0$ obtained from the
minimization of the ground-state energy~(\ref{eq:gs_energy})
while the vertical line shows the critical temperature~(\ref{eq:tc}).
}
\end{figure}

As the grand-canonical potential~(\ref{eq:free_energy3}) has one minimum at $\delta=0$
for high temperatures and two minima at $\pm \delta \neq 0$ at zero temperature,
there is a phase transition at a (mean-field) critical temperature  $T_{\text{P}}$.
One can easily determine numerically the value of the order parameter that
minimizes~(\ref{eq:free_energy3}) for a given temperature $T$. The result
for $\lambda=0.18$  is shown
in Fig~\ref{fig:delta}.
Clearly, there is a continuous transition from the high-temperature uniform metallic phase 
to the low-temperature dimerized insulating phase at a finite critical temperature 
$k_{\text{B}}T_{\text{P}} \approx 0.104 t_0$.

\begin{figure}
\includegraphics[width=0.9\columnwidth]{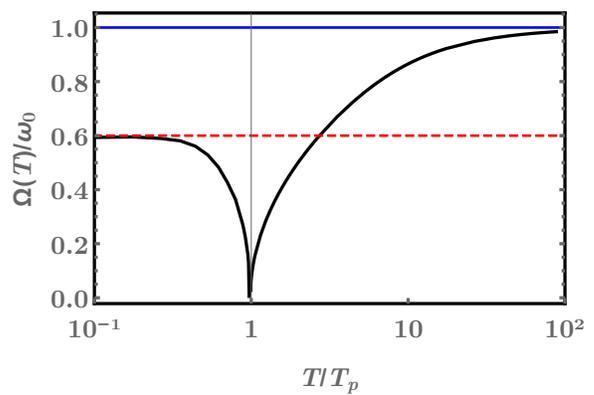}
\caption{\label{fig:omega} 
Renormalized phonon frequency $\Omega/\omega_0$ for the amplitude mode 
[see Eq.~(\ref{eq:omega})] 
as a function of temperature at half filling with an electron-phonon coupling $\lambda = 0.18$.  
The horizontal dashed and solid lines show the zero-temperature frequency~(\ref{eq:omega0})
and bare phonon frequency, respectively.
The vertical line shows the critical temperature $T_{\text{P}}$, see Eq.~(\ref{eq:tc}).
}
\end{figure}

Figure~\ref{fig:omega} shows that the renormalized phonon frequency~(\ref{eq:omega})
calculated numerically from the second derivative of (\ref{eq:free_energy3}) at its minimum.
We see that the phonon mode becomes completely soft at the critical temperature 
already found for the order parameter in Fig.~\ref{fig:delta}.
This vanishing of the amplitude mode frequency is the signature of the Kohn anomaly in the phonon spectrum 
around the wave number $Q=2k_{\text{F}}$~\cite{heeg88,baeriswyl92,solyom3,schu78,horo78,tuti91,Gruener00} 
in a mean-field description of the Peierls transition.

From the necessary condition for a minimum
\begin{equation}
\label{eq:extrema}
\frac{d\phi}{d\delta}=0
\end{equation}
one can deduce the self-consistency equation for solutions $\delta \neq 0$
\begin{equation}
\label{eq:sce}
\frac{1-\delta^2}{2\lambda} 2t_0 = \int_{\vert \Delta \vert}^{2t_0} d\varepsilon
\sqrt{\frac{4t^2_0-\varepsilon^2}{\varepsilon^2 - \Delta^2}}
\frac{\sinh(\beta \varepsilon)}{1+\cosh(\beta \varepsilon)}.
\end{equation}
If we approximate the spectrum~(\ref{eq:dispersion}) by linear dispersions
for  $\vert \varepsilon \vert \gg \vert \Delta \vert$   and treat $2t_0$ as
the ultraviolet cutoff,
we can approximate $\sqrt{4t^2_0-\varepsilon^2} \approx 2t_0$ in the above integral.
In the weak-coupling regime ($\lambda \ll 1 \Rightarrow \delta_0 \ll 1$)
we can then solve the self-consistency equation for $T=0$ and for
 $\delta\rightarrow 0 \ ( \Leftrightarrow T\rightarrow T_{\text{P}})$.
In the ground state ($T=0$) we obtain  
\begin{equation}
\label{eq:delta0prime}
\delta_0^{\prime} = 2 \exp\left (-\frac{1}{2\lambda}\right ),
\end{equation}
which deviates from the correct result~(\ref{eq:delta0}) only by a constant factor
$e/2$.
This factor [like the one due to the SSH hopping term~(\ref{eq:hopping}), see Appendix~\ref{app:hopping}]  
is only a minor problem because of the one-to-one correspondence between $\delta_0,\delta_0^{\prime}$ and the single relevant model parameter $\lambda$.
Thus one can deduce relations between observables and $\delta_0^{\prime}$ using the above approximation
and then substitute the correct value $\delta_0$ to obtain a quantitatively accurate result. 
This is especially well illustrated by the following determination of the critical temperature.

At the critical temperature one can insert $\delta=0$ in the self-consistency equation and thus find 
\begin{equation}
\label{eq:tc}
k_{\text{B}} T^{\prime}_{\text{P}} = \frac{2t_0e^{\gamma}}{\pi} \delta^{\prime}_0 \approx 1.134  \delta^{\prime}_0 t_0
\end{equation}
with the Euler constant $\gamma \approx 0.577$.
This leads to the well-known relation between the (mean-field) critical temperature and 
the zero-temperature electronic gap~(\ref{eq:gap})
~\cite{Gruener00}
\begin{equation}
\label{eq:tc2}
k_B T^{\prime}_{c} =  \frac{e^{\gamma}}{2\pi} E^{\prime}_{\text{g}} \approx \frac{E^{\prime}_{\text{g}}}{3.53} .
\end{equation}
We see in Figs.~\ref{fig:delta} and~\ref{fig:omega} that the relation~(\ref{eq:tc}) 
yields an excellent approximation of the critical temperature for an electron-phonon
coupling as large as $\lambda=0.18$. However, we have to substitute 
the correct value the zero-temperature order parameter~(\ref{eq:delta0}), 
in lieu of the approximate value~(\ref{eq:delta0prime}), in
the relation~(\ref{eq:tc}) to obtain this quantitative agreement.
For stronger couplings deviations become significant. For instance,
for $\lambda = 0.37$ we obtain $\delta_0 \approx  0.446$
which corresponds to $k_{\text{B}}T_{\text{P}} \approx 0.505 t_0$ 
according to~(\ref{eq:tc}) but we find numerically that 
$k_{\text{B}}T_{\text{P}} \approx 0.455 t_0$.

\section{Results for fixed chemical potential \label{sec:gc}}
 
We now turn to the general case of a fixed chemical potential $\mu < 0$.
(As mentioned earlier the case $\mu > 0$ yields similar results.)
For high temperatures ($k_{\text{B}} T \gg t_0$) the grand-canonical potential~(\ref{eq:free_energy2}) 
can be approximated by
\begin{equation}
\phi = \frac{t_0}{\pi \lambda} \delta^2 -  \mu - k_{\text{B}} T \ln(2+2\cosh(\beta\mu))  . 
\end{equation}
Therefore, the uniform metallic configuration $(\delta =0)$ 
is the only stable phase at high temperatures for any value of $\mu$.

\subsection{Ground-state results \label{sec:gs}}

The ground-state results (i.e., for $T \rightarrow 0$) are more interesting.
The grand-canonical potential~(\ref{eq:free_energy}) yields in the thermodynamic limit
\begin{eqnarray}
\label{eq:gs_energy2}
\phi & = & \frac{t_0}{\pi \lambda} \delta^2 - \mu \rho - \frac{4t_0}{\pi} E\left (\frac{\pi}{2} \rho, 1-\delta^2
\right)
\end{eqnarray}
where $E(\phi,m)$ is the incomplete elliptic integral of the second kind.
The electronic density is $\rho=1$ when the chemical potential lies in the 
Peierls gap ($0 \geq \mu \geq -\vert \Delta \vert$), 
\begin{equation}
\rho = \frac{2}{\pi} \arcsin \left (\sqrt{\frac{4t_0^2 - \mu^2}{4t_0^2 - \Delta^2}} \right )
\in (0,1)
\end{equation}
when it is within the valence band  ($-\vert \Delta \vert > \mu > -2t_0$), 
and $\rho=0$ when it lies below the valence band ($\mu \leq -2t_0$). 
These three different relative values of
the chemical potential $\mu$ are illustrated in Fig.~\ref{fig:dispersion}. 

We can now examine the zero-temperature phases as a function of the chemical potential.
 We note that Eq.~(\ref{eq:gs_energy2}) for $\mu \geq -\vert \Delta \vert$ is equal to the ground-state
energy at half filling~(\ref{eq:gs_energy}) up to a constant shift $-\mu$. Consequently, $\phi$ varies with $\delta$
as at half filling as long as the chemical potential lies in the electronic band gap.
Thus we know that for $\mu > -\Delta_0 = -2\delta_0 t_0$ 
there are two (possibly local) minima at $\delta = \pm \delta_0$ (as at half filling)
and no other extrema for larger values of $\vert \delta \vert$.
Similarly to the half-filling case~(\ref{eq:condensation}) the minima are given at weak coupling by 
\begin{equation}
\label{eq:gs_energy3}
\phi_{\text{min}} = -\frac{4 t_0}{\pi} -\mu -  \frac{t_0}{\pi} \delta_0^2.
\end{equation}
Additionally, we can conclude  that  there is no extremum
for $- \vert \Delta \vert < \mu < -\Delta_0$. 

For $\mu  <  -\vert \Delta \vert$ but $ \vert\mu\vert \ll 2t_0$ we can expand 
Eq.~(\ref{eq:gs_energy2}) up to the second order in $\delta$. Using the weak-coupling result for
half filling~(\ref{eq:delta0}) we find
\begin{equation}
\label{eq:expansion2}
\phi = -\frac{4 t_0}{\pi} -\mu - \frac{\mu^2}{2\pi t_0} + \frac{2t_0}{\pi} 
\ln \left (\frac{2\vert \mu\vert}{\Delta_0} \right) 
\delta^2.
\end{equation}
For $\mu > -\Delta_0/2$ the coefficient of the quadratic term is negative
and thus there is a local maximum at $\delta=0$ as for half filling.
For $\mu < -\Delta_0/2$, however, the coefficient of $\delta^2$ is positive. Thus
there is a (possibly local) minimum at $\delta=0$.
Comparing the energies~(\ref{eq:gs_energy3}) of the minima at $\delta = \pm \delta_0$
and~(\ref{eq:expansion2}) for the minimum at $\delta=0$ we see that
the dimerized configuration has a lower energy for $\mu > - \sqrt{2} \delta_0 t_0 = - \Delta_0/\sqrt{2}$
while the uniform configuration becomes stable when $\mu$ drops below this value.

From the necessary condition~(\ref{eq:extrema}) one can deduce 
a self-consistency equation for the extrema of~(\ref{eq:gs_energy2})
at finite $\delta$. Assuming again a linear electronic dispersion with a cutoff $2t_0$ for $\vert \varepsilon \vert \gg \vert \Delta \vert$, we obtain solutions
that agree with the above analysis of the potential~(\ref{eq:gs_energy2}). 
Moreover, the absence of any solution of the self-consistency equation 
for
$\mu < -\Delta_0$ shows that the dimerized state is unstable in this regime.
See the appendix~\ref{app:sce} for more details.

In summary, the weak-coupling analysis reveals the existence of four phases
at zero temperature. 
In the first phase for $0 \geq \mu > -\Delta_0/2$, which includes the half-filled band case,
the dimerized configuration is stable and the uniform configuration is unstable. 
In the second phase $-\Delta_0/2 > \mu > -\Delta_0/\sqrt{2}$ the dimerized configuration is still stable but 
the uniform configuration is now metastable.
In the third phase for $ -\Delta_0/\sqrt{2} > \mu > -\Delta_0$ 
the dimerized configuration becomes metastable while the uniform configuration is stable.  
Finally, in the fourth phase for $-\Delta_0  > \mu$ 
the uniform configuration is still stable and the dimerized configuration is unstable.
Moreover, the order parameter of the (stable or metastable) dimerized configuration has the same value
(\ref{eq:delta0}) as the one found at half filling down to $\mu = -\Delta_0$.
Thus the value of the zero-temperature order parameter as a function of the chemical potential ($\mu \leq 0$)
is
\begin{equation}
\delta(T=0,\mu) = \left \{
\begin{aligned}[l,c,l]
\pm \delta_0 & \ \ \text{for} & \mu > -\Delta_0/\sqrt{2}  \\
0 & \ \ \text{for} & \mu < -\Delta_0/\sqrt{2} 
\end{aligned}
\right . .
\end{equation}

\begin{figure}
\includegraphics[width=0.9\columnwidth]{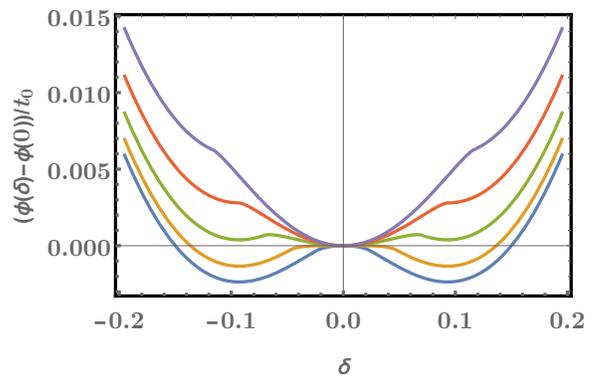}
\caption{\label{fig:potential2}
Grand-canonical potential~(\ref{eq:free_energy2}) at zero-temperature [see Eq.~(\ref{eq:gs_energy2})]
as a function of the order parameter $\delta$ 
for $\lambda=0.18$ and several values of the chemical potential from $\mu=-0.25 \Delta_0$ (bottom curve) to 
$\mu=-1.25 \Delta_0$ (top curve) in steps of $-0.25\Delta_0$.
}
\end{figure}

\begin{figure}
\includegraphics[width=0.9\columnwidth]{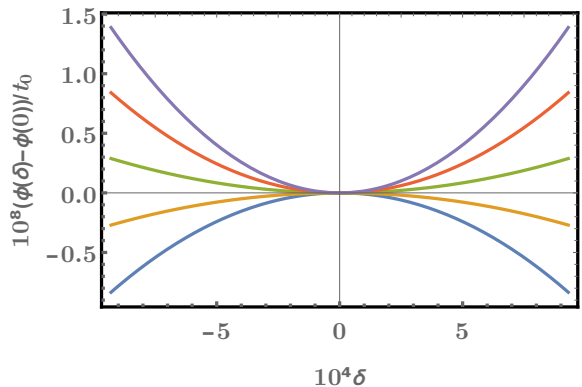}
\caption{\label{fig:potential2b}
Grand-canonical potential~(\ref{eq:free_energy2}) at zero-temperature [see Eq.~(\ref{eq:gs_energy2})]
as a function of the order parameter $\delta$ around $\delta=0$
for $\lambda=0.18$ and several values of the chemical potential from $\mu=-0.485 \Delta_0$ (bottom curve) to 
$\mu=-0.505 \Delta_0$ (top curve) in steps of $-0.005\Delta_0$.
}
\end{figure}

\begin{figure}[b]
\includegraphics[width=0.9\columnwidth]{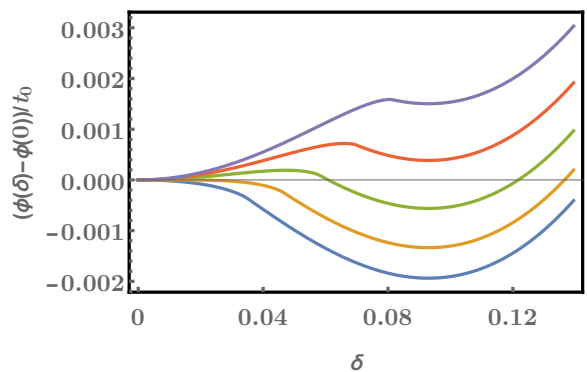}
\caption{\label{fig:potential2c}
Grand-canonical potential~(\ref{eq:free_energy2}) at zero-temperature [see Eq.~(\ref{eq:gs_energy2})]
as a function of the order parameter $\delta \geq 0$ 
for $\lambda=0.18$ and several values of the chemical potential from $\mu=-0.375 \Delta_0$ (bottom curve) to 
$\mu=-0.875 \Delta_0$ (top curve) in steps of $-0.125\Delta_0$.
}
\end{figure}

The potential~(\ref{eq:gs_energy2}) can easily be calculated numerically as a function of $\delta$
for a given chemical potential.
The results are shown in Figs.~\ref{fig:potential2}, \ref{fig:potential2b}, and~\ref{fig:potential2c}
for the finite electron-phonon coupling $\lambda=0.18$.
Figure~\ref{fig:potential2} shows that 
for $\mu=-\Delta_0/4$ the shape of $\phi$ still resembles the double-well potential shown in Fig.~\ref{fig:potential}
for half filling.
We see clearly that 
the minima corresponding to the dimerized configuration are progressively raised (relative
to the potential of the uniform configuration $\delta=0$) as $\mu$ is lowered
until they become metastable (see the curve for $\mu=-0.75\delta_0$),
and are finally suppressed (see the curve for $\mu=-1.25\delta_0$).

Figure~\ref{fig:potential2b} shows the behavior of $\phi$ around $\delta=0$ in more details.
We clearly see that the local maximum at $\delta=0$ becomes a minimum when the chemical
potential decreases from $-0.49\Delta_0$ to $-0.495\Delta_0$. The small deviation
from the weak-coupling boundary value $-\Delta_0/2$ is due to the finite value of $\lambda$.
The deviation grows larger with $\lambda$, for instance the boundary value reaches 
$\mu=-0.43 \Delta_0$ for $\lambda=0.37$.

Finally, the existence of metastable states is illustrated in Fig.~\ref{fig:potential2c}.
Clearly, we observe an absolute minimum of $\phi$ at finite $\delta$ 
and a local minimum at  $\delta=0$ for $\mu=-0.625\Delta_0$
while for $\mu=-0.750\Delta_0$  there is an absolute minimum
at $\delta=0$ and a local minimum at finite $\delta$.
This agrees with the weak-coupling critical value $\mu = -\Delta_0/\sqrt{2} \approx -0.71 \Delta_0$
for the boundary between dimerized and uniform phases.
Moreover, we see in Fig.~\ref{fig:potential2c} that the order parameter of the (stable or metastable) dimerized state
(i.e., the position of the minimum with $\delta \neq 0$) does not vary with $\mu$ in agreement with the weak-coupling prediction.

Therefore, the numerical results for the finite electron-phonon coupling $\lambda=0.18$
in Figs.~~\ref{fig:potential2}, \ref{fig:potential2b}, and~\ref{fig:potential2c} confirm
the weak-coupling analysis. In particular, we have found a dimerized phase 
with a metastable uniform state for $-\Delta_0/2 > \mu > -\Delta_0/\sqrt{2}$ at zero temperature.

\subsection{Finite-temperature phase diagram \label{sec:pd}}

We now turn our attention to the still unexplored case $T > 0$ and $\mu < 0$.
From the necessary condition for extrema~(\ref{eq:extrema}) one can deduce 
the self-consistency equation for the minima and maxima of the grand-canonical potential~(\ref{eq:free_energy2})
at finite $\delta$ 
\begin{equation}
\label{eq:sce2}
\frac{1-\delta^2}{2\lambda} 2t_0 = \int_{\vert \Delta \vert}^{2t_0} d\varepsilon
\sqrt{\frac{4t^2_0-\varepsilon^2}{\varepsilon^2 - \Delta^2}}
\frac{\sinh(\beta \varepsilon)}{\cosh(\beta \mu) +\cosh(\beta \varepsilon)}.
\end{equation}
The solutions $\delta(T,\mu)$ can be easily computed numerically
but it is difficult to determine the phase boundary of a continuous transition.
However, if $\delta(T,\mu)$ vanishes continuously, the critical values  of $T$ and $\mu$
must satisfy the self-consistency equation
\begin{equation}
\label{eq:sce3}
\frac{2t_0}{2\lambda} = \int_{0}^{2t_0} d\varepsilon
\frac{\sqrt{4t^2_0-\varepsilon^2}}{\varepsilon}
\frac{\sinh(\beta \varepsilon)}{\cosh(\beta \mu) +\cosh(\beta \varepsilon)}.
\end{equation}

\begin{figure}
\includegraphics[width=0.99\columnwidth]{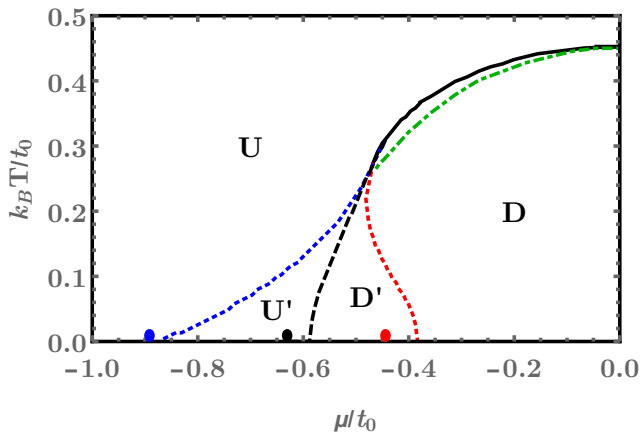}
\caption{\label{fig:phasediagram}
Phase diagram ($\mu,T$) of the SSH model in the mean-field approximation 
for an electron-phonon coupling $\lambda=0.37$. 
The black solid and dashed lines show the phase boundary $T_{\text{P}}(\mu)$ between
the dimerized phase (labeled D or D$^{\prime}$) and the uniform phase (labeled U or U$^{\prime}$).
The uniform configuration is metastable in the sector labeled D$^{\prime}$ while the dimerized
state is metastable in the sector labeled U$^{\prime}$.
The solid line indicates the continuous phase transition between the sectors D and U while 
the dashed line indicates the first-order transition between the sectors  D$^{\prime}$  and U$^{\prime}$
with metastable configurations.
The dot-dashed line indicates the position of the first-order metal-insulator transition
within the dimerized phase.
The circles show the zero-temperature boundaries predicted by the weak-coupling analysis.
}
\end{figure}

The ($T,\mu$) phase diagram for the SSH model in the mean-field approximation
can be determined numerically using these two equations.
The result is shown in Fig.~\ref{fig:phasediagram} for 
the electron-phonon coupling $\lambda=0.37$ [corresponding
to $\delta_0 = 0.446$,  $\Delta_0 = 0.891 t_0$, and 
$k_{\text{B}}T_{\text{P}}(\mu=0)  \approx 0.455 t_0$].
Results are qualitatively similar for weaker $\lambda$.
We see that the grand-canonical phase diagram is much richer than for the Peierls
transition at fixed band filling (for instance,  for $\mu=0$).
It consists  of two main phases, the dimerized phase (noted D or 
D$^{\prime}$) and the uniform phase (labeled U or U$^{\prime}$).
We denote $T_{\text{P}}(\mu)$, or equivalently $\mu_{\text{P}}(T)$,
the boundary line between these two phases.
Each phase is made of two sectors. 
In the main sectors (D or U) only one state (dimerized or uniform)
is stable. In the smaller sectors D$^{\prime}$ and U$^{\prime}$
the other state is metastable.
We denote  $T_{\text{D}}(\mu)$ [or $\mu_{\text{D}}(T)$] the boundary line
between the sectors U and  U$^{\prime}$, where the metastable dimerized configurations
vanish. Similarly, $T_{\text{U}}(\mu)$ [or $\mu_{\text{U}}(T)$] denotes 
the boundary line
between the sectors D and  D$^{\prime}$, where the metastable uniform configuration
vanishes.
The dimerized and uniform phases coexist only on the boundary between 
the sectors D$^{\prime}$ and U$^{\prime}$.
The coexistence terminates at  a critical point 
($T_{\text{c}},\mu_{\text{c}}$).
We could not estimate this point analytically but from the solution
of the self-consistency equation~(\ref{eq:sce2})
we obtain the position $k_{\text{B}}T_{\text{c}} \approx 0.25 t_0 $ and $\mu_{\text{c}} \approx -0.48 t_0 $
for $\lambda=0.37$.

The transition between the sectors D and U is continuous while
it is first-order between the sectors with metastable configurations
D$^{\prime}$ and U$^{\prime}$.   
Note that the boundary of the sector $D$ at finite $T$ and $\mu$ is
given by the solutions of the self-consistency equation~(\ref{eq:sce3})
while the boundary of the sector $U$  is determined by
the solution of the self-consistency equation~(\ref{eq:sce2}) with the 
highest temperature for a given chemical potential.
The boundary between the sectors D$^{\prime}$ and U$^{\prime}$
is given by the solutions of~(\ref{eq:sce2}) with the same potential~(\ref{eq:free_energy2})
as the uniform configuration $\delta=0$.

Figure~\ref{fig:phasediagram} reveals that
the sector boundaries at low temperature 
[$\mu_{\text{D}}(T\rightarrow 0) \approx -0.87 t_0$, 
$\mu_{\text{P}}(T\rightarrow 0)  \approx -0.586 t_0$, 
and $\mu_{\text{U}}(T\rightarrow 0) \approx -0.385 t_0$]
are close to the ground-state
results of Sec.~\ref{sec:gs}
[$\mu_{\text{D}}(T=0)= - \Delta_0 \approx -0.891 t_0 $, 
$\mu_{\text{P}}(T=0)= - \Delta_0 /\sqrt{2} \approx -0.630 t_0$, and $\mu_{\text{U}}(T=0) = -\Delta_0/2 \approx -0.446 t_0$],
although deviations are clearly visible for the electron-phonon coupling
$\lambda=0.37$ used here.
For smaller electron-phonon couplings, such as $\lambda=0.18$,
 the phase diagram determined numerically
using Eqs.~(\ref{eq:sce2}) and~(\ref{eq:sce3}) agree quantitatively
with the ground-state boundaries given in Sec.~\ref{sec:gs}.

In the phase diagram in Fig.~\ref{fig:phasediagram} we also see that the system moves rapidly through the four sectors
if one changes  the chemical potential at fixed but low temperature.
In particular, the system undergoes a first-order transition at the critical
value $\mu_{\text{P}}(T)$.
This could explain the sensitivity of the In/Si(111) system  to the chemical doping
of the substrate~\cite{zhan14,shim09,mori10,schm11},
 which corresponds to changing the chemical potential in our model.

\begin{figure}
\includegraphics[width=0.9\columnwidth]{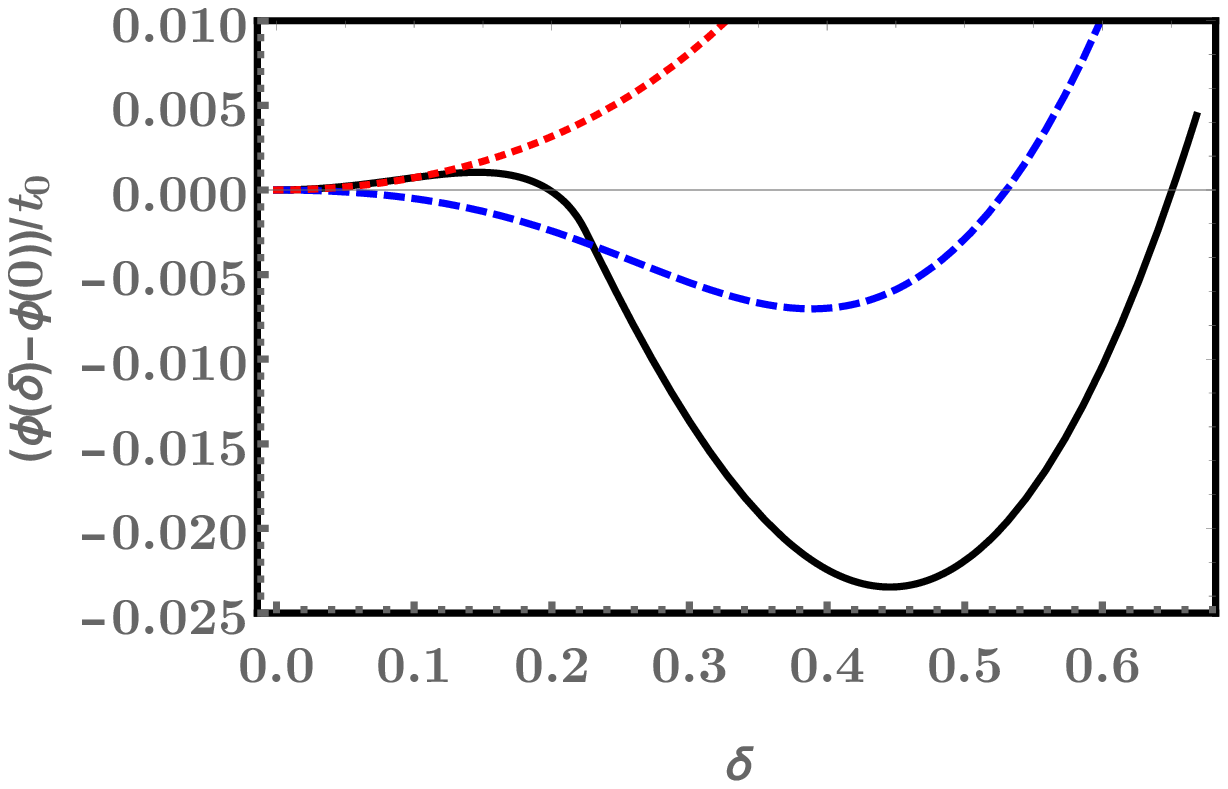}
\caption{\label{fig:potentialr2}
Grand-canonical potential~(\ref{eq:free_energy2}) 
as a function of the order parameter $\delta \geq 0$ for
$\lambda = 0.37$, $\mu= -0.5 \Delta_0 \approx- 0.446 t_0$ 
and several temperatures $k_{\text{B}}T=0.005 \Delta_0 \approx 0.00446 t_0$ (black solid line),
$k_{\text{B}}T=0.225 \Delta_0 \approx  0.201 t_0$ (blue dashed line), and
$k_{\text{B}}T=0.45 \Delta_0 \approx 0.401 t_0$ (red dotted line).
}
\end{figure}

\begin{figure}
\includegraphics[width=0.9\columnwidth]{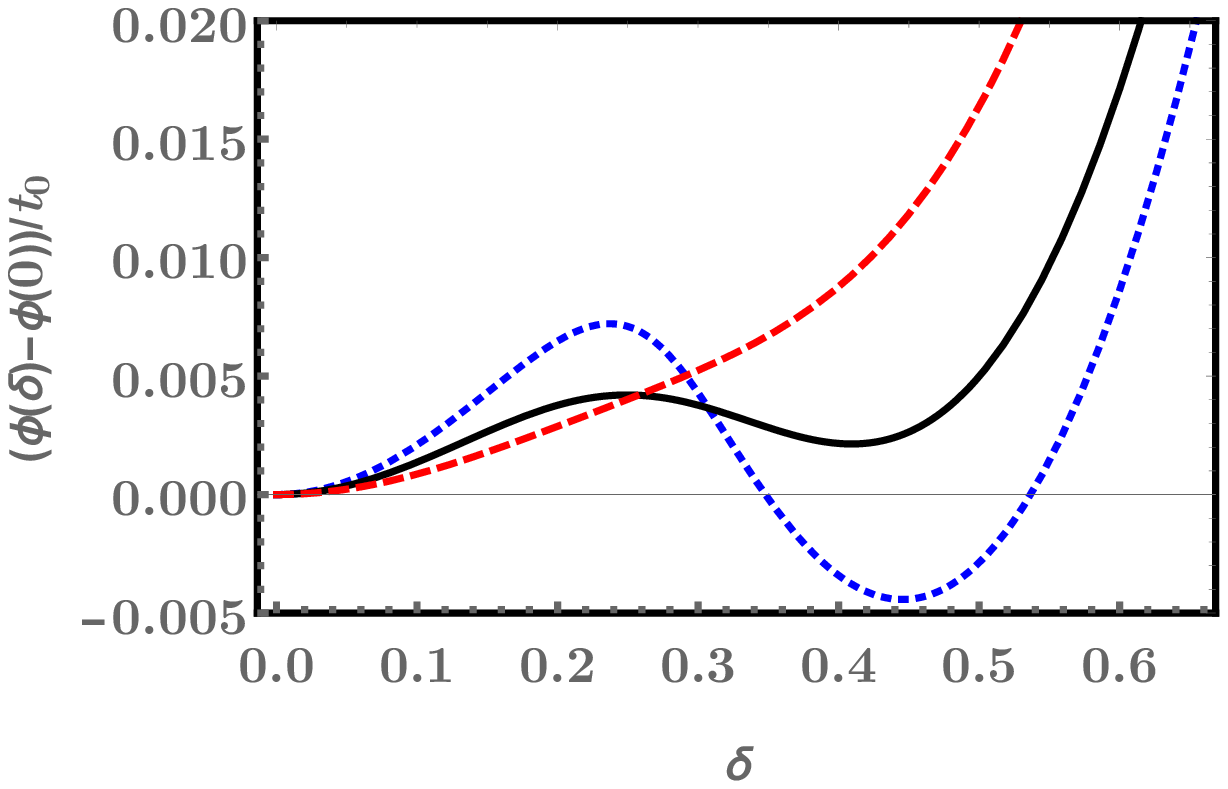}
\caption{\label{fig:potentialr3}
Grand-canonical potential~(\ref{eq:free_energy2}) 
as a function of the order parameter $\delta \geq 0$ for
$\lambda = 0.37$, $\mu= -0.625 \Delta_0 \approx-0.557 t_0$ 
and several temperatures $k_{\text{B}}T=0.05 \Delta_0 \approx 0.0446 t_0$ (blue dotted line),
$k_{\text{B}}T=0.15 \Delta_0 \approx 0.134 t_0$ (black solid line), and
$k_{\text{B}}T=0.25 \Delta_0 \approx 0.223 t_0$ (red dashed line).
}
\end{figure}

\begin{figure}
\includegraphics[width=0.9\columnwidth]{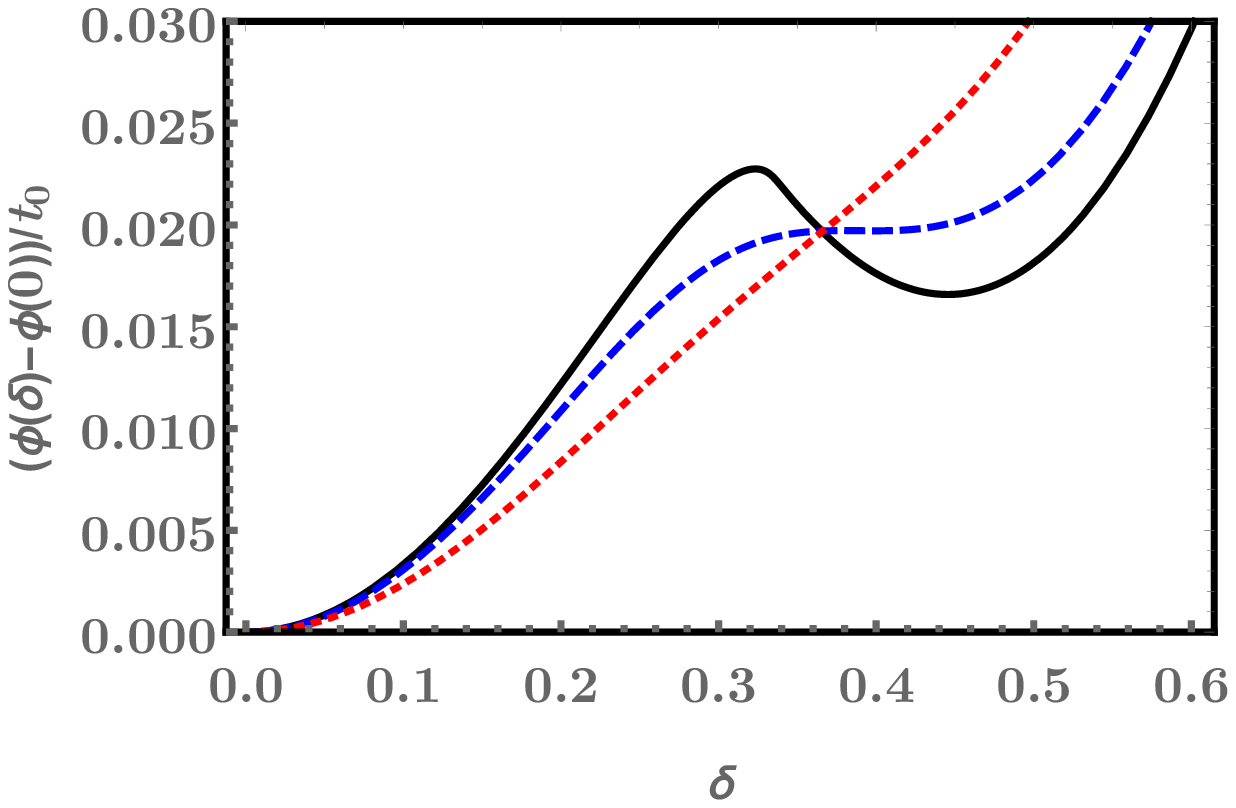}
\caption{\label{fig:potentialr4}
Grand-canonical potential~(\ref{eq:free_energy2}) 
as a function of the order parameter $\delta \geq 0$ for
$\lambda = 0.37$, $\mu= -0.75 \Delta_0 \approx - 0.668 t_0$ 
and several temperatures $k_{\text{B}}T=0.005 \Delta_0 \approx 0.00446 t_0$ (black solid line),
$k_{\text{B}}T=0.1 \Delta_0 \approx  0.0891 t_0$ (blue dashed line), and
$k_{\text{B}}T=0.2 \Delta_0 \approx 0.178 t_0$ (red dotted line).
}
\end{figure}

To understand this phase diagram better we discuss 
the evolution of the system when temperature is raised at a fixed chemical potential
in more details.
There are three unusual scenarios that can happen depending on the value 
of the chemical potential: $\mu_{\text{U}}(T=0) > \mu > \mu_{\text{c}}$,
$\mu_{\text{c}}> \mu > \mu_{\text{P}}(T=0)$,
and $\mu_{\text{P}}(T=0) > \mu > \mu_{\text{D}}(T=0)$.
The variations of the grand-canonical potential are illustrated
in Figs.~\ref{fig:potentialr2}, ~\ref{fig:potentialr3}, and~\ref{fig:potentialr4}
for these three cases.
For comparison, the variation  of the free energy with  temperature
can be seen  in Fig.~\ref{fig:potential} for  the usual continuous Peierls transition at
 a fixed band filling. This scenario remains qualitatively valid
 as long as the uniform configuration is unstable at low temperature,
 explicitly for $\mu > \mu_{\text{U}}(T=0)$ with
 $\mu_{\text{U}}(T=0) \approx -0.385 t_0$ for $\lambda=0.37$ in the phase diagram
 in Fig.~\ref{fig:phasediagram} .
 
We consider first $\mu= -0.5 \Delta_0 \approx-0.446 t_0$, which lies between  
$\mu_{\text{U}}(T=0) \approx -0.385 t_0$ and $\mu_{\text{c}} \approx -0.48 t_0$.
We see in the phase diagram in Fig.~\ref{fig:phasediagram}  that
for this value of $\mu$ the system is dimerized with a metastable uniform phase
at low temperature but moves
from the sector D$^{\prime}$ to the sector D with increasing $T$
because the metastable uniform configuration vanishes for temperatures higher than
$k_{\text{B}}T_{\text{U}}(\mu) \approx 0.11 t_0$. Then the system
undergoes a continuous transition to the uniform phase U
at a critical temperature $k_{\text{B}}T_{\text{P}}(\mu) \approx 0.307 t_0$.
The grand-canonical potential~(\ref{eq:free_energy2}) shown in Fig.~\ref{fig:potentialr2} changes  accordingly,
from a function with absolute minima for finite $\delta$ and a local minimum 
at $\delta=0$ for a temperature lower than $T_{\text{U}}(\mu)$, to 
the usual double-well shape for a temperature between  $T_{\text{U}}(\mu)$
and $T_{\text{P}}(\mu)$, and finally to a single-well shape for a temperature
higher than $T_{\text{P}}(\mu)$.

Second, we examine the case of a chemical potential $\mu=-0.625 \Delta_0 \approx-0.557 t_0$,  which lies
between $\mu_{\text{c}} \approx -0.48 t_0$ and $\mu_{\text{P}}(T=0) \approx -0.586 t_0$.
We see in the phase diagram in Fig.~\ref{fig:phasediagram}  that
 the system is again dimerized with a metastable uniform phase
at low temperature for this value of $\mu$. However, it now undergoes
a first-order transition from the sector D$^{\prime}$ to the sector U$^{\prime}$
at the critical temperature $k_{\text{B}}T_{\text{P}}(\mu) \approx 0.102 t_0$.
As the temperature increases further the system moves into the sector U
because the metastable dimerized configurations vanish above 
$k_{\text{B}}T_{\text{D}}(\mu) \approx 0.165 t_0$.
The grand-canonical potential~(\ref{eq:free_energy2}) in Fig.~\ref{fig:potentialr3} changes  accordingly,
from a function with absolute minima for finite $\delta$ and a local minimum 
at $\delta=0$ for a temperature lower than $T_{\text{P}}(\mu)$ to 
a function with an absolute minimum at $\delta=0$ and local minima for finite $\delta$
between  $T_{\text{P}}(\mu)$ and $T_{\text{D}}(\mu)$, and finally exhibits the usual single-well shape for a temperature
higher than $T_{\text{D}}(\mu)$.

Third, we discuss the case of $\mu= -0.75 \Delta_0 \approx -0.668 t_0$, which lies between
$\mu_{\text{P}}(T=0) \approx  -0.586 t_0$ and 
$\mu_{\text{D}}(T=0) \approx -0.87 t_0$.
In that case the system remains 
within the uniform phase for all temperatures but moves from the sector U$^{\prime}$ to the sector U
as the temperature is raised because the 
metastable dimerized state vanishes for 
temperatures higher than $k_{\text{B}}T_{\text{D}}(\mu) \approx 0.087 t_0$.
Accordingly,  Fig.~\ref{fig:potentialr4} shows that the grand-canonical potential~(\ref{eq:free_energy2})
changes from a function with an absolute minimum at $\delta=0$ and  local minima at finite $\delta$
for a temperature below $T_{\text{D}}(\mu)$
 to the usual single-well shape for high temperatures.

The above discussion completely describes the structural transition, which is uniquely determined by
the order parameter  through Eq.~(\ref{eq:Delta}).
The order parameter also determines  the size of the electronic gap $E_{\text{g}}$
that is open by the Peierls lattice distortion through Eq.~(\ref{eq:gap}).
The vanishing of $\delta(T,\mu)$ implies the closing of this gap
and thus
 in a Peierls transition at fixed band filling the metal-insulator transition occurs simultaneously 
to the structural transition and is also continuous.
At a fixed chemical potential, however, the metal-insulator transition occurs when the chemical potential
reaches the upper edge of the valence band, i.e. when
$\mu = E_{\text{g}}/2 = - \Delta(T,\mu)$.
Thus this transition 
can take place at a lower temperature than the structural transition 
and in this case it is first order, as the gap jumps from $0$ to $E_{\text{g}}$ at this point.
We have found that this scenario occurs in the sector D of the dimerized phase
for all chemical potentials $0 > \mu > \mu_{\text{c}}$.
The critical temperature for the metal-insulator transition is shown in  Fig.~\ref{fig:phasediagram}
and is in fact slightly lower than  $T_{\text{P}}(\mu)$ for a given $\mu$.
As the structural transition is first order for $ \mu < \mu_{\text{c}}$, we conclude
that the metal-insulator transition is first order for all $0 > \mu > \mu_{\text{P}}(T=0) $
in the dimerized SSH model in the mean-field approximation.

\subsection{First-order Peierls transition}

With a view to understanding the first-order transition in In/Si(111),
the most  interesting part of the phase diagram is
the region with a chemical potential 
between $\mu_{\text{c}} $ and $\mu_{\text{P}}(T=0)$.
As discussed above, this leads to a first-order transition from the dimerized
insulating phase at low temperature to a metallic uniform phase 
at high temperature with a metastable uniform configuration
below the critical temperature.
This agrees with  the experimental observations for
 In/Si(111)~\cite{hatt11,wall12,schm12,klas14,zhan14,Jeckelmann16,spei16,hatt17}.
 Therefore, we examine the physical properties of the SSH model in that regime
using the parameter  $\lambda=0.37$ and $\mu =-0.625 \Delta_0 \approx -0.557 t_0$
corresponding to the second case discussed above in the previous section.

\begin{figure}
\includegraphics[width=0.9\columnwidth]{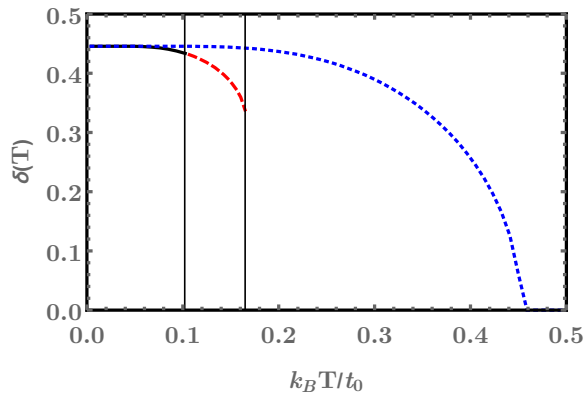}
\caption{\label{fig:delta2}
Order parameter $\delta$ as a function of temperature for $\lambda = 0.37$.
The vertical lines indicate the critical temperature $k_{\text{B}}T_{\text{P}}(\mu) \approx 0.102 t_0$
and $k_{\text{B}}T_{\text{D}}(\mu) \approx 0.165 t_0$ for $\mu=-0.625 \Delta_0 \approx -0.557 t_0$.
The solid black and red dashed lines show $\delta$ for the thermodynamically stable dimerized phase 
[$T < T_{\text{P}}(\mu)$]
and the metastable dimerized state [$T_{\text{P}}(\mu) < T < T_{\text{D}}(\mu)$]
at this value of $\mu$, respectively.
The blue dotted line shows $\delta(T)$ at half filling ($\mu=0$).
}
\end{figure}

As discussed in Sec.~\ref{sec:model}  the order parameter $\delta(T,\mu)=\Delta(T,\mu)/(2t_0)$
determines both the lattice deformation $u$ through Eq.~(\ref{eq:Delta}) 
and the electronic gap $E_{\text{g}}$ through Eq.~(\ref{eq:gap}).
Figure~\ref{fig:delta2} shows the order parameter $\delta$ as a function of temperature
through the first-order transition.
We see that $\delta$ diminishes first progressively as the temperature rises from $T=0$ to
$k_{\text{B}}T_{\text{D}}(\mu) \approx 0.165 t_0$, above which the dimerized state becomes 
unstable and  $\delta$ drops to zero. However, the dimerized state is thermodynamically
stable only up to the lower critical temperature $k_{\text{B}}T_{\text{P}}(\mu) \approx 0.102 t_0$ and thus
the first-order structural and metal-insulator transition takes place already at this temperature.
This is in strong contrast to the continuous Peierls transition at half filling ($\mu=0$),
which is also shown in Fig.~\ref{fig:delta2}.

\begin{figure}
\includegraphics[width=0.9\columnwidth]{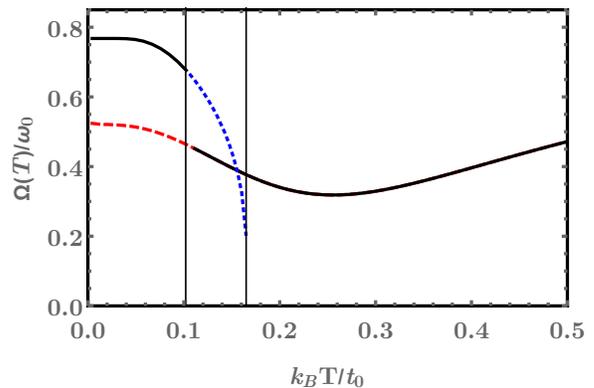}
\caption{\label{fig:omega2}
Renormalized phonon frequency $\Omega/\omega_0$ [Eq.~(\ref{eq:omega})] as a function of temperature for $\lambda = 0.37$
and $\mu=-0.625 \Delta_0 \approx -0.557 t_0$.
The vertical lines indicate the critical temperature $k_{\text{B}}T_{\text{P}}(\mu) \approx 0.102 t_0$
and $k_{\text{B}}T_{\text{D}}(\mu) \approx 0.165 t_0$.
The solid black line shows $\Omega$ in the thermodynamically stable phases with a jump
at $T = T_{\text{P}}(\mu)$. The red dashed and blue dotted lines show the frequency for the 
metastable uniform configuration below $T_{\text{P}}(\mu)$ and the metastable dimerized configurations for
$T_{\text{P}}(\mu) < T < T_{\text{D}}(\mu)$, respectively.
}
\end{figure}

Another interesting physical  quantity is the frequency of the Raman-active amplitude oscillations
of the order parameter~\cite{schu78,horo78,tuti91,Gruener00}, which can be measured experimentally for In/Si(111)~\cite{Jeckelmann16,spei16}.
Figure~\ref{fig:omega2} shows the renormalized phonon frequency $\Omega(T)/\omega_0$  calculated from Eq.~(\ref{eq:omega})
as a function of temperature through the first-order transition.
We see that the frequency of the thermodynamically stable state diminishes progressively
as the temperature rises from to $T=0$ to $k_{\text{B}}T_{\text{P}}(\mu) \approx 0.102 t_0$.
At this critical temperature the phonon frequency jumps to a lower value and then changes
again smoothly as the temperature increases further. 
[For $\mu$ close to $\mu_{\text{c}}$  the frequency can also jumps up at $T_{\text{P}}(\mu)$.]
This jump of $\Omega(T)$ corresponds to the abrupt transition
from oscillations around the dimerized configurations to oscillations around the uniform configuration.
This variation  of the phonon frequency with the temperature is in strong contrast to the  
complete phonon softening  found in a continuous Peierls transition at fixed band filling 
and shown in Fig.~\ref{fig:omega}.

In Fig.~\ref{fig:omega2} we also show the phonon frequencies for oscillations around 
the metastable configurations. They prolong smoothly the curves $\Omega(T)/\omega_0$
obtained for the stable configurations.
Note that amplitude oscillations around the dimerized configurations have the wave number $Q=2k_{\text{F}}=\pi/a$
but they appear at $Q=0$ in experiments  below the critical temperature $T_{\text{P}}(\mu)$ because of the folding
of the Brillouin zone due to the lattice dimerization.
Oscillations around the uniform configuration  always correspond to $Q=0$.

\section{Conclusion \label{sec:conclusion}}

We have investigated the grand-canonical Peierls transition in the SSH model in the mean-field approximation.
We have found that the phase diagram is much richer than for the Peierls theory at fixed band filling.
This could explain the sensitivity of  the transition in In/Si(111) to chemical doping of the substrate~\cite{zhan14,shim09,mori10,schm11}. Notably, we have found a first-order Peierls transition
from the insulating dimerized phase to the metallic uniform phase
when temperature is raised at fixed chemical potential. Moreover, the uniform phase 
remains metastable below the critical temperature. These findings agree with 
 experimental evidence and first-principles simulations for
 In/Si(111)~\cite{hatt11,wall12,schm12,klas14,zhan14,Jeckelmann16,spei16,hatt17}.
 Therefore,  we think that this grand-canonical Peierls theory is 
 the appropriate basis for describing the quasi-one-dimensional physics realized in the In/Si(111) system.

The SSH model is a minimal model for a commensurate Peierls transition
and we have investigated its phase diagram using a basic mean-field approach.
Thus we briefly discuss some effects that we have neglected and 
possible extensions of the present work.

Our approach neglects spatial fluctuations of the lattice distortion $u_j$.
Away from half filling ($\mu \neq 0$)  the lattice  is probably unstable
with respect to incommensurate Peierls distortions, such as 
\begin{equation}
u_j = \tilde{u} \cos(Q j a + \phi)
\end{equation}
with appropriate amplitude $\tilde{u}$ and  wave number $Q$ as well as an arbitrary phase $\phi$.
(The wave number would be  $Q=2k_F$ in the Peierls theory at fixed band filling~\cite{Gruener00}.)
However, these incommensurate distortions have a much lower critical temperature than the
dimerization. Moreover,  the coupling between atomic wires and the periodic substrate lattice
should render them even less energetically stable.
Thus we do not expect them to play any significant role as long as $\mu > -\Delta_0$,
but they could become relevant in the sector U of the uniform phase at very low temperature.

A more interesting effect of spatial fluctuations of the order parameter is that 
the dimerized phase could be unstable with respect to the formation of domain walls between 
both dimerized configurations  (so-called solitons). 
The theoretical modeling of solitons is a key feature of the SSH model~\cite{su79,su80,heeg88,baeriswyl92}
and  the existence and properties of solitons in the In/Si(111) system are still debated~\cite{mori04,zhan11a,yeom11,zhan11b,kim12,cheo15,lee19}.
Therefore, it would be interesting to extend the grand-canonical Peierls theory to describe
solitons.

Besides spatial fluctuations, thermal and quantum lattice fluctuations are important in low-dimensional systems~\cite{Gruener00,baeriswyl04,baeriswyl92,horo75,lee73}.
It is well-known that a spontaneous symmetry breaking such as the dimerization
cannot occur at finite temperature in a one-dimensional systems with finite range interactions
because of thermal fluctuations.
However, it is also well-established that true finite-temperature 
phase transitions may take place in quasi-one-dimensional systems
made of a higher dimensional array of (weakly) coupled chains because the fluctuations are 
suppressed by the ordering perpendicular to the chains.~\cite{horo75,baeriswyl04,solyom2}
In atomic wires on substrates,  both this coupling between wires and 
the coupling of the wires to the substrate suppress fluctuations
and thus will allow for a Peierls transition at finite temperature.
Nevertheless, it will be necessary to 
extend the present work to include inter-wire coupling combined with thermal and quantum fluctuations
to study the phase transition and the system properties
in the critical regime more accurately.
Their effects can be studied in SSH-like models using sophisticated numerical methods such as quantum Monte Carlo simulations~\cite{webe16,hohe18,webe18}.
However, we think that our results for the first-order transition and the  metastable states
away from the critical point $(T_{\text{c}},\mu_{\text{c}})$ will remain
qualitatively valid when they are taken into account. 
The main combined effects of inter-wire coupling  and fluctuations
will be to reduce the various temperatures calculated within the mean-field semi-classical 
approach, as it was found for the Peierls theory at fixed band filling~\cite{Gruener00,lee73,horo75,tuti91}.

Electronic correlation effects are also neglected in the SSH model as they do not play
 a determinant role in the Peierls theory.  
 Typically, one assumes that  the Coulomb repulsion between electrons 
only renormalizes the model parameters such as the electron-phonon coupling~\cite{heeg88}.
However, it is known that they can be important for a correct description of important aspects
 of the Peierls physics~\cite{baeriswyl92, barford2013,tima17}, such as 
the insulating dimerized phase with domain walls away from half filling (soliton lattice)~\cite{jeck94,jeck98b}
or the possible Luttinger liquid properties of the metallic phase just above the critical temperature~\cite{Oncel08}.
 Therefore, it could be interesting to study the grand-canonical Peierls
 transition in generalizations of the SSH model
including  the electron-electron interaction explicitly.

In summary,  we think that various aspects neglected in the present work 
(solitons, thermal and quantum fluctuations, inter-wire coupling, electronic correlations) 
should be investigated in the future 
to achieve a more realistic and accurate description of the grand-canonical Peierls transition in atomic wires deposited on 
semiconducting substrates. Nevertheless, 
we are confident that our main findings, the occurrence of a first-order Peierls transition and 
the metastability of the uniform state,  will remain relevant.

\begin{acknowledgments}
This work was done
as part of the Research Unit \textit{Metallic nanowires on the atomic scale: Electronic
and vibrational coupling in real world systems} (FOR1700) 
of the German Research Foundation (DFG) and was supported by
grants Nos.~JE~261/1-2.
\end{acknowledgments}

\appendix

\section{Hopping term \label{app:hopping}}

The SSH linear approximation for the hopping term~(\ref{eq:hopping}) is not consistent with the harmonic
approximation for the lattice elastic energy~(\ref{eq:potential}) because second-order contributions
of the lattice displacements $u_j$ 
to the electronic energy are thus neglected.  
As an example, consider a 
(more accurate)  exponential dependence of the hopping terms on the distance variation between 
sites~\cite{long59,sale60} 
\begin{equation}
\label{eq:hopping2}
t( u_{j+1} - u_j) = t_0 \exp \left [ -\alpha (u_{j+1} - u_j)/t_0  \right ].
\end{equation}
Expanding up to second order yields 
\begin{equation}
\label{eq:hopping3}
t( u_{j+1} - u_j) = t_0 - \alpha (u_{j+1} - u_j) + \frac{\alpha^2}{2t_0} (u_{j+1} - u_j)^2.
\end{equation}
This agrees with the SSH hopping term~(\ref{eq:hopping}) up to first order in the displacements $u_j$ 
but the second order term is completely neglected in the SSH model.
Using the second-order expansion~(\ref{eq:hopping3})  and the dimerized configuration~(\ref{eq:mf}),
we recover  the electronic dispersion~(\ref{eq:dispersion})
but with 
\begin{equation}
\label{eq:hopping4}
\tilde{t}_0 = t_0 \left ( 1 + \frac{1}{2} \delta^2  \right )
\end{equation}
substituted for $t_0$.
An analysis similar to the one carried out in Sec.~\ref{sec:half}
leads to similar results, in particular the  zero-temperature order parameter
is given by 
\begin{equation}
\label{eq:delta02}
\tilde{\delta}_0 =  \frac{\tilde{\Delta}_0}{2t_0} =  4 \exp\left ( -\frac{1}{2\lambda} \right)
\end{equation}
in the weak-coupling limit.
This differs by a constant factor $e^{-1}$ from the SSH result~(\ref{eq:delta0}).
Therefore, disregarding the second-order term in~(\ref{eq:hopping3}) leads
to a quantitatively incorrect result. 

This factor does not play any role in the qualitative investigation of the SSH model properties, however,
because  there is  a one-to-one relation between the order parameter 
and the only model parameter $\lambda$.
As
 the renormalization of the bare hopping term~(\ref{eq:hopping4}) is otherwise
negligible in the weak-coupling limit, 
this allows us to express
all  observables directly  as a  function of  $\tilde{\delta}_0$ rather than $\lambda$
and to verify that the (weak-coupling) model properties depend on $\delta_0$ [for the hopping term~(\ref{eq:hopping})]  
exactly as on $\tilde{\delta}_0$ [for the hopping term~(\ref{eq:hopping3})].
Therefore, the  different prefactors in Eqs.~(\ref{eq:delta0}) and~(\ref{eq:delta02})
[as well as (\ref{eq:delta0prime})] must only be taken into account
when comparing to numerical simulation of the SSH model~\cite{Jeckelmann16}, first-principles simulations,
or experimental results.

\section{Zero-temperature self-consistency equation away from half filling \label{app:sce}}

From the general self-consistency equation~(\ref{eq:sce2}) one obtain 
for $T\rightarrow 0$
\begin{equation}
\label{eq:sce4}
\frac{1-\delta^2}{2\lambda} 2t_0 = \int_{\Lambda}^{2t_0} d\varepsilon
\sqrt{\frac{4t^2_0-\varepsilon^2}{\varepsilon^2 - \Delta^2}}
\end{equation}
where $\Lambda$ is the largest of $\vert \Delta \vert$ and $\vert \mu \vert$.
 Assuming again a linear electronic dispersion with a cutoff $2t_0$ for $\vert \varepsilon \vert \gg \vert \Delta \vert$
 and $\vert \delta \vert  \ll 1$, 
 we obtain the same self-consistency equation for $\Lambda = \vert \Delta \vert$ as for half filling and thus the same result~(\ref{eq:delta0prime}) for the solution $\delta_0^{\prime}$. Thus there are
 two (possibly local) minima at $\pm \delta_0^{\prime}$ for  $0 \geq \mu \geq -\Delta_0^{\prime} = 2\delta_0^{\prime} t_0$ and no extrema with $\vert \Delta \vert >  \vert \mu \vert > \Delta_0^{\prime}$.
 For $\Lambda = \vert \mu \vert$ the self-consistency equation becomes
 \begin{equation}
 \ln \left ( \frac{\Delta_0^{\prime}}{\vert \Delta \vert} \right ) = \text{acosh}\left (\frac{\vert \mu \vert }{\vert \Delta \vert }\right ) \geq 0.
 \end{equation}
 Obviously, there is no solution with $\vert \mu \vert > \vert \Delta \vert > \Delta_0^{\prime}$.
 Consequently, there are no extrema with $\delta \neq 0$ for any $\mu <  -\Delta_0^{\prime}$.
 We see that the above equation possesses solutions $\pm \Delta^{\prime} \neq 0$ starting from 
 $\Delta^{\prime}=0$
 for $\mu = - \Delta_0^{\prime}/2$ and increasing continuously up to
 $\Delta^{\prime} = \Delta_0^{\prime}$ for $\mu = - \Delta_0^{\prime}$.
 One can check numerically that these solutions of the self-consistency equation
 correspond to local maxima.
 Thus this analysis confirms the existence of simultaneous minima
 for the uniform and dimerized configurations when 
 $- \Delta_0^{\prime}/2 > \mu > -\Delta_0^{\prime}$
 in agreement with the discussion of Eq.~(\ref{eq:gs_energy2}) in Sec~\ref{sec:gs}.
Moreover, the absence of any solution of the self-consistency equation 
for $\mu < -\Delta_0^{\prime}$ shows that the dimerized state is unstable in this regime.

\vfill 

\bibliographystyle{biblev1}
\bibliography{mybibliography}{}

\end{document}